\input{psfig}
\documentstyle [12pt,twoside,epsf]{article}

\newcommand{\bq}{\begin{equation}}
\newcommand{\ba}{\begin{eqnarray}}
\newcommand{\eq}{\end{equation}}
\newcommand{\ea}{\end{eqnarray}}

\def\b{\beta}

\def\f{\phi}

\def\l{\lambda}

\def\p{\pi}

\def\F{\Phi}

\def\J{\Psi}
\def\L{\Lambda}

\def\bo{{\raise.15ex\hbox{\large$\Box$}}}
\def\bob{{\lower.2ex\hbox{\large$\Box$}}}
\def\pa{\partial}

\catcode`@=11
\def\underline#1{\relax\ifmmode\@@underline#1\else
        $\@@underline{\hbox{#1}}$\relax\fi}
\catcode`@=12

\begin{document}

\hfill{LA-UR-94-3816}

\centerline{\large{\bf {$\F^4$ Kinks: Statistical
Mechanics}}\footnote{This paper is based on a presentation by the author
at NEEDS '94, Los Alamos, NM (September 1994)}}

\vspace{1.5cm}

\centerline{Salman Habib}

\vspace{1.5cm}

\centerline{\em T-6, Theoretical Astrophysics}
\centerline{\em and}
\centerline{\em T-8, Elementary Particles and Field Theory}
\centerline{\em Los Alamos National Laboratory}
\centerline{\em Los Alamos, NM 87545}

\vspace{2.5cm}

\centerline{\bf Abstract}

Some recent investigations of the thermal equilibrium properties of
kinks in a $1+1$-dimensional, classical $\Phi^4$ field theory are
reviewed.  The distribution function, kink density, correlation
function, and certain thermodynamic quantities were studied both
theoretically and via large scale simulations. A simple double
Gaussian variational approach within the transfer operator formalism
was shown to give good results in the intermediate temperature range
where the dilute gas theory is known to fail.

\vfill
\noindent e-mail:\\
\noindent habib@predator.lanl.gov\\
\newpage

\section{Introduction}

The statistical mechanics of nonlinear coherent structures such as
solitons and solitary waves has been a subject of study for quite some
time \cite{SS}. More recent interest has been fueled by applications
not only in condensed matter physics \cite{CM}\cite{BKT}, but also by
potential applications in particle physics (sphalerons) \cite{SR} and
cosmology (domain walls, baryogenesis) \cite{KT}. 

A particularly simple model that displays behavior representative of a
large class of soliton-bearing systems is the ``double-well'' $\F^4$
scalar field theory in $1+1$-dimensions. This theory admits exact
solitary wave solutions (``kinks'') and our purpose here is to
describe the equilibrium statistical mechanics of these objects as
studied recently in Refs. \cite{AH}\cite{AHK}. 

The Lagrangian for this theory is
\begin{eqnarray}
L= {1\over 2}\left(\pa_t\Phi\right)^2
   -{1\over 2}\left(\pa_x\Phi\right)^2
   +{1\over 2}m^2\Phi^2-{1\over 4}\Lambda\Phi^4.                \label{1}
\end{eqnarray}
The equations of motion that follow from (\ref{1}) admit static kink
solutions (centered at $x=x_0$) 
\bq
\F_k={m\over\L}\tanh\left({m\over\sqrt{2}}(x-x_0)\right).  \label{kink}
\eq
The total energy of an isolated kink is $E_k=\sqrt{8/9}m^3/\L$ and the
negative of a kink, also a solution, is called an antikink.

In  numerical work it is customary to use the dimensionless form of
this theory, given by the transformations: $\Phi\to a\phi$, $x\to
x/m$, and $t\to t/m$, where $a^2=m^2/\lambda$. The equation of motion
then becomes 
\begin{eqnarray}
\pa^2_{tt}\phi=\pa^2_{xx}\phi -\phi\left(\phi^2-1\right).          \label{2}
\end{eqnarray}

The statistical mechanics of kinks in this system has been studied by
two approaches. In the first, and phenomenological, approach one assumes
that the kinks and the fluctuations (``phonons'') about the asymptotic
field minima may be treated as weakly interacting elementary
excitations. Provided the kink density is low, the canonical partition
function can then be found by standard methods
\cite{SS}\cite{KS}\cite{CKBT}. Alternatively, as shown by Krumhansl
and Schrieffer (KS) \cite{KS}, building on earlier work of Sears,
Scalapino, and Ferrell \cite{SSF}, it is possible to calculate the
partition function, in principle exactly, by exploiting a transfer
operator technique. KS showed that in the low temperature (``dilute
gas'') limit the partition function naturally factorizes into two
contributions both having counterparts in the phenomenological theory;
a tunneling term which they were able to identify with the kink
contribution, and the remainder which they identified as linearized
phonons. The ideas of KS were further refined and extended to a wider
class of systems by Currie {\em et al} \cite{CKBT}. In particular,
interactions of kinks with linearized phonons were considered, leading
to substantial corrections to the results of KS.

A key result of these efforts is the prediction that the spatial
density of kinks
\begin{equation}
n_k\propto\sqrt{E_k\beta}\exp(-E_k\beta),                    \label{dgnum}
\end{equation}
where $E_k$ is the kink energy ($E_k=\sqrt{8/9}$ for the dimensionless
form of the theory) and $\beta$, the inverse temperature (for the
dimensionless case, $\beta\to\beta/(a^3\sqrt{\lambda}$)). A related
prediction is that at low temperatures the field correlation length
$\lambda$ defined by 
\begin{equation}
\langle\phi(0)\phi(x)\rangle\sim\exp\left(-x/\lambda\right) \label{cl}
\end{equation}
has an exponential temperature dependence \cite{CKBT},
\bq
\lambda={1\over 4}\sqrt{{\p\over 3}}
{1\over\sqrt{E_k\b}}\exp(E_k\beta).      \label{CL}
\eq

Computer simulations to verify these results date back to the work of
Koehler {\em et al} \cite{KBKS} who found only a qualitative
agreement. More recent investigations \cite{GR}\cite{BD}\cite{AFG} led
to more detailed comparisons; however, significant discrepancies were
reported. This led to theoretical speculation \cite{FM}\cite{BF2}
regarding possible corrections to the dilute gas theory of kinks. It
was speculated that these discrepancies could be due to finite size
effects and phonon dressing of the bare kink energy (breather
contributions to the free energy could also be significant
\cite{SSB}). However, these simulations were not carried out at low 
enough temperatures and the authors interpreted their results in terms
of WKB (dilute gas) formulas that, it turns out, were simply not valid
over the range of temperatures studied \cite{AH}\cite{AHK}. In
Refs. \cite{AH}\cite{AHK}, by going to low enough temperatures it was
shown that the standard dilute gas results were indeed valid.
Furthermore, an earlier claim of substantial phonon dressing even at
these temperaures \cite{KBKS} was shown not to be correct.

The equilibrium statistical mechanics of kinks in the $\Phi^4$ model
(\ref{1}) was investigated by implementing a Langevin code on a
massively parallel computer \cite{AH}. The key idea is to supplement the
equation of motion (\ref{2}) with noise and viscosity terms obeying an
appropriate fluctuation-dissipation theorem so that the system is
driven to thermal equilibrium at the desired temperature \cite{ScS}. To
understand the numerical results in the high and intermediate
temperature region not susceptible to a dilute gas analysis, a double
Gaussian wave function approximation for the quantum
mechanical problem which results from applying the transfer operator
method was used. Not only is this method accurate but it also suggests a
natural decomposition of nonlinear phonon and kink degrees of freedom
in the intermediate temperature regime \cite{AHK}. 

The main results of these investigations are: (1) the dilute gas
predictions for the kink density and the correlation length are in
good agreement with the simulations below a certain (theoretically
estimable) temperature, (2) above this temperature the double Gaussian
results for the kink number and the correlation length agree with the
simulations, (3) kinks are found to ``disappear'' into the thermal
phonon background above a characteristic temperature, in good
agreement with theoretical prediction, (4) the double Gaussian
approximation accurately describes the classical single point field
distribution function at high and intermediate temperatures where the
dilute gas (WKB) approximation breaks down, and (5) the internal
energy and the specific heat calculated in the double Gaussian
approximation show an interesting energy sharing process between kinks
and nonlinear phonons in an intermediate-temperature range below the
characteristic temperature at which kinks appear: a peak in the
specific heat in this temperature range is shown to be due essentially
to kinks. 

\section{Numerical Results}

The Langevin equation for the dimensionless theory is
\begin{eqnarray}
\pa^2_{tt}\phi=\pa^2_{xx}\phi-\eta\pa_t\phi-\phi(1-\phi^2) + F(x,t).
\end{eqnarray}
To guarantee an approach to equilibrium, the Gaussian, white noise $F$
and the viscosity $\eta$ are related via the fluctuation-dissipation
theorem:
\begin{eqnarray}
\langle F (x,t) F (x',t') \rangle = 2\eta\bar{\beta}^{-1}\delta (x-x')
\delta (t-t').
\end{eqnarray}
Numerical simulations were performed on lattices with 16384 sites and
the Langevin equation was solved numerically using standard methods
\cite{AH}. The system size was $1$ to $2$ orders of magnitude larger
than that in most previous simulations. Large system sizes are
necessary to get acceptable statistics at low temperatures and to
avoid finite size effects. Systems were evolved from a random initial
condition to equilibrium.

\begin{figure}
\centerline{\psfig{figure=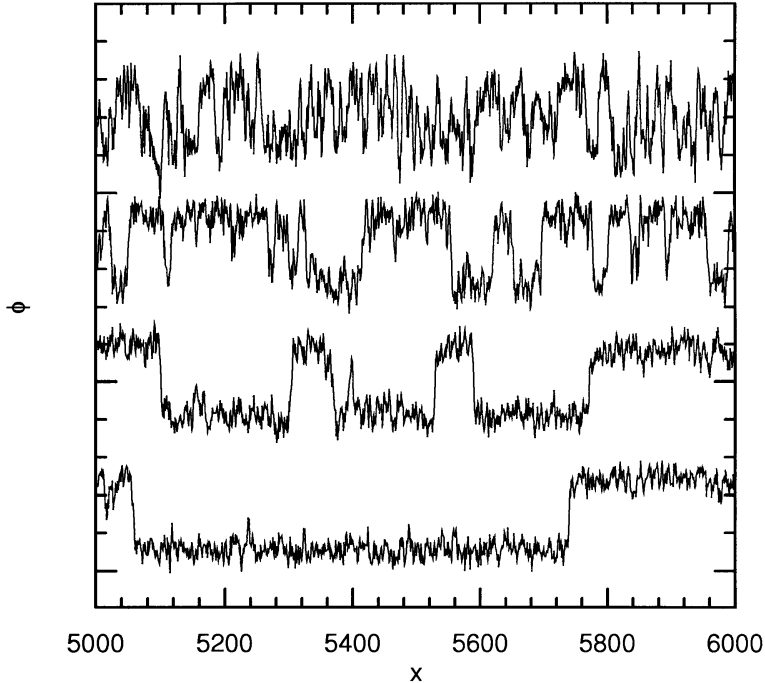,height=9cm,width=10cm}}
\caption[Figure 1]{\small{Sample field configurations, from top to
bottom, at $\bar{\beta}=2,4,5.5,8$. Only a 1000 lattice unit sample
of the total lattice size of 16384 is shown.}}
\end{figure}

To compute the kink number an operational way to identify kinks is
needed even though the exact kink solution is available theoretically.
Several possible definitions were examined, all of which relied on a
knowledge of the kink size.  From the classical solution for a kink
centered at $x_0$, $\phi=tanh((x-x_0)/\sqrt{2})$, the kink scale $L_k$
was approximately 8 lattice units in the simulations. Raw kink
configurations are shown in Fig. 1. At low temperatures ($\bar{\beta}
>5$), kinks may be identified easily, however at higher temperatures
this is clearly not the case.

The simplest thing to do is to count the number of zero-crossings of
the field, since one may argue that these are the ``tunneling events''
which correspond to kinks. However, at higher temperatures there are
zero-crossings due to thermal noise (phonons), and counting all
zero-crossings would lead to a gross overestimation of the number of
kinks. At high temperatures it is not possible to distinguish
unambiguously between kinks and nonlinear phonons (the overcounting
problem occurs even at intermediate temperatures where the kinks are
more or less distinct). A possible solution is to use a smoothed field
by either averaging or block-spinning the actual field configuration
over a length of the order of the kink scale. The latter approach was
taken in previous simulations \cite{GR}\cite{BD}\cite{AFG}. This
solution is not without flaws either, as rapid fluctuations can still
appear as kinks. In Refs. \cite{AH}\cite{AHK} the following method was
used: at a particular time find all zero-crossings. To test the
legitimacy of a given zero-crossing check for zero-crossings one kink
scale (8 lattice units) to its right and to its left. If no
zero-crossings are found, count it as a kink; otherwise not.

\begin{figure}
\centerline{\psfig{figure=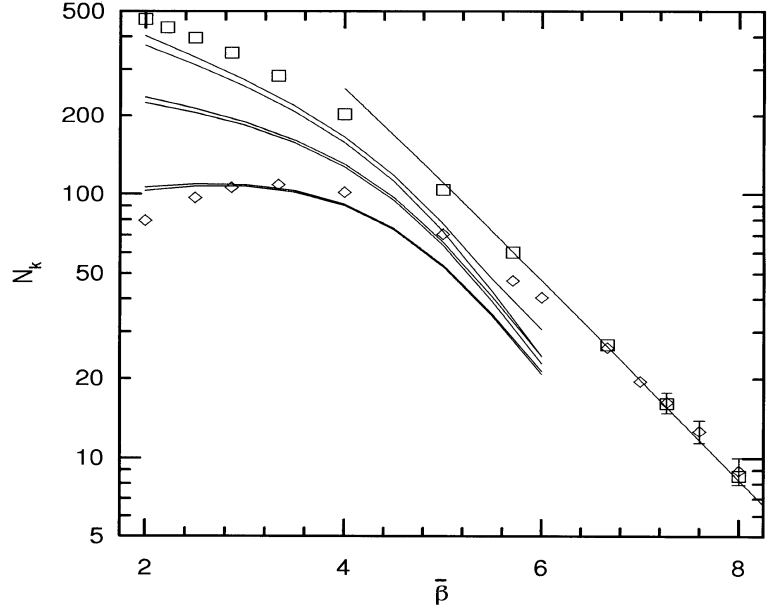,height=8cm,width=10cm}}
\caption[Figure 2]{\small{Total number of kinks as a function of
$\bar{\beta}$. Squares denote counts with a smoothed field (averaging
length of 8 lattice units) definition of kinks, diamonds for the
zero-crossing counting method discussed in Sec. 2, and the solid
beginning at $\bar{\beta}=4$ is a fit to the WKB prediction
(\ref{dgnum}). Also shown are three predictions from the double
Gaussian theory (for details, see Ref. \cite{AHK}).}}
\end{figure}

The number of kinks is plotted against $\bar{\beta}$ in Fig. 2. Above
$\bar{\beta}\sim 6$, the two methods for counting kinks
agree. Moreover, in this (low temperature) range, the dilute gas
expression for the kink number (\ref{dgnum}) is in excellent agreement
with the data. At elevated temperatures, there is a clear disagreement
between the two methods of counting kinks. In this temperature regime
the number of kinks computed with the smoothing method depends
strongly on the smoothing scale. For $\bar{\beta} < 6$, the number of
kinks cannot be extracted with any confidence from the smoothing
method.

The correlation length $\lambda$ is plotted against $\bar{\beta}$ in
Fig. 3. For $\bar{\beta} > 6$, the WKB prediction (\ref{cl}) holds,
whereas for $\beta < 5$, where the wave function overlaps are not
negligible, the double Gaussian approximation is valid. (Fortunately,
there are no ambiguities here with regard to measurements at higher
temperatures.) 

\begin{figure}
\centerline{\psfig{figure=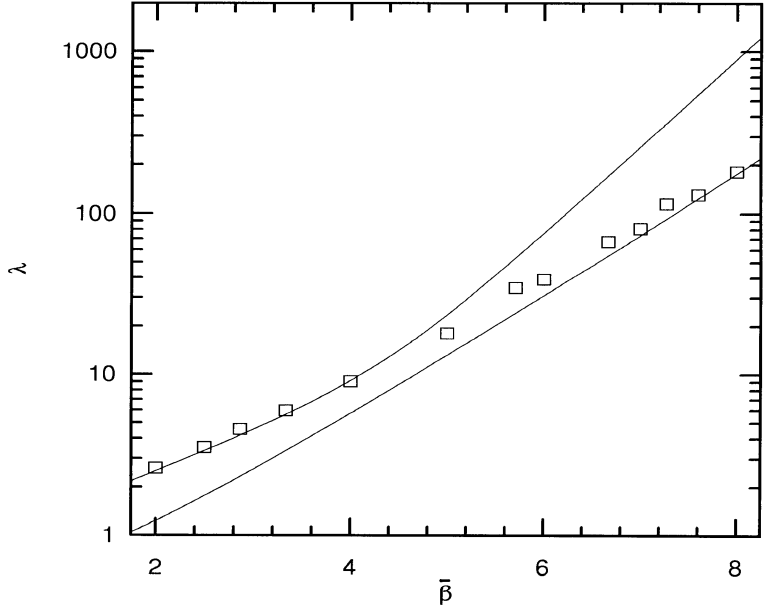,height=8cm,width=10cm}}
\caption[Figure 3]{\small{The correlation length $\l$ as a function of
$\bar{\beta}$. The two solid lines are the theoretical predictions
(top, double Gaussian and bottom, WKB). The crossover of the ranges of
validity of the two theories occurs at $\bar{\beta}\sim 5$.}}
\end{figure}

\section{Transfer Operator and Double Gaussian Approximation}

The canonical partition function for the Lagrangian (\ref{1}) is
given by the functional integral
\begin{eqnarray}
Z = \int D\phi D\pi~\exp\left(-\beta H(\phi,\pi)\right) \label{Z}
\end{eqnarray}
where $\pi$ is the canonical momentum of the field and $H$, the field
Hamiltonian. The transfer operator technique \cite{SSF} reduces the
calculation of the partition function in the thermodynamic limit to
simply finding the ground state energy of the double well quantum
Hamiltonian (here written for the dimensionless case),
\begin{equation}
H_Q={1\over 2}p^2 - {1\over 2\beta^2}{\bar{\phi}}^2 + {1\over
4\beta^4}{\bar{\phi}}^4               \label{ham}
\end{equation}
where ${\bar{\phi}}=\beta\phi$. At low temperatures the two wells are
widely separated and the ground state energy is given by the
oscillator ground state energy for one of the wells minus the
tunnel-splitting term, usually calculated by WKB methods. The dilute
gas or WKB approximation for the kink number is valid when the
tunnel-splitting is small enough such that only the first two energy
eigenstates are necessary to estimate the ground state energy of the
Hamiltonian (\ref{ham})\cite{BK}. At higher temperatures where kinks
still exist, higher energy states cannot be ignored. Since kinks are
associated with tunneling, one expects them to vanish when the ground
state energy is higher than the classical barrier height: this
intuition is confirmed by the simulations \cite{AH}\cite{AHK}.

One can compare the simulations of the kink system with numerical
solutions for the energy eigenvalues of the Hamiltonian
$H_Q$. Instead, a different approach may be taken by implementing a
double Gaussian variational method (see \cite{AHK} for details) which
is an order of magnitude more accurate than the simple Gaussian
approximation \cite{GA} for this problem and correctly accounts for
the reduction of energy due to overlap terms in the wave functions of
the two wells, at least for moderate to large overlaps. Three
qualitatively different regimes exist: (1) all the energy eigenvalues
lie above the classical barrier, (2) the ground state energy lies
below the classical barrier height, and (3) the energy difference
between the ground and first excited state becomes negligible in
comparison with the energy difference between the ground and the
second excited state (this occurs for $\beta > 6$). The simulations
\cite{AH} confirm the theoretical expectations that there are no kinks
in region (1), that there are kinks, but that the dilute gas
approximation is invalid in region (2), and finally, that the dilute
gas approximation is accurate in region (3) (a regime unexplored in
detail by previous simulations).

The classical single point field distribution function $P[{\phi}]$ was
measured directly in the simulations. For the analogous quantum
mechanical problem this is just the square of the ground state wave
function $\Psi_0$. Results from the simulations and the double
Gaussian approximation are compared in Fig. 4. The presence of kinks
implies a double peak in $P[\phi]$ \cite{ADB}(the converse is false)
while a single peak at the origin means that kinks and thermal phonons
can no longer be distinguished. From the simulations such a transition
occurs at $\beta\simeq 1.7$, in agreement with the theoretical
calculation of when $\Psi_0^{~2}$ goes over from a double to single
peaked distribution. As expected, this is also the temperature
($\beta=1.734$) where the ground state energy crosses the classical
barrier height (a discussion of various methods to determine the
characteristic temperature is given in Ref. \cite{ARB}). The double
peaks in the distribution function move inward from the classical
minimum as the temperature increases (this is clearly seen in
Fig. 4). Physically this can be understood as nonlinear phonon
corrections due to the fact that near each minimum, the potential is
not symmetric under reflection around the minimum. The double Gaussian
approximation underestimates the tunneling or overlap contribution at
low temperatures ($\sim \exp(-\sqrt{2}\bar{\beta})$ for double
Gaussian versus $\sim \exp(-\sqrt{8/9}\bar{\beta})$ for WKB). This can
be seen already at $\bar{\beta}=4$ in Fig. 4. 

\begin{figure}
\centerline{\psfig{figure=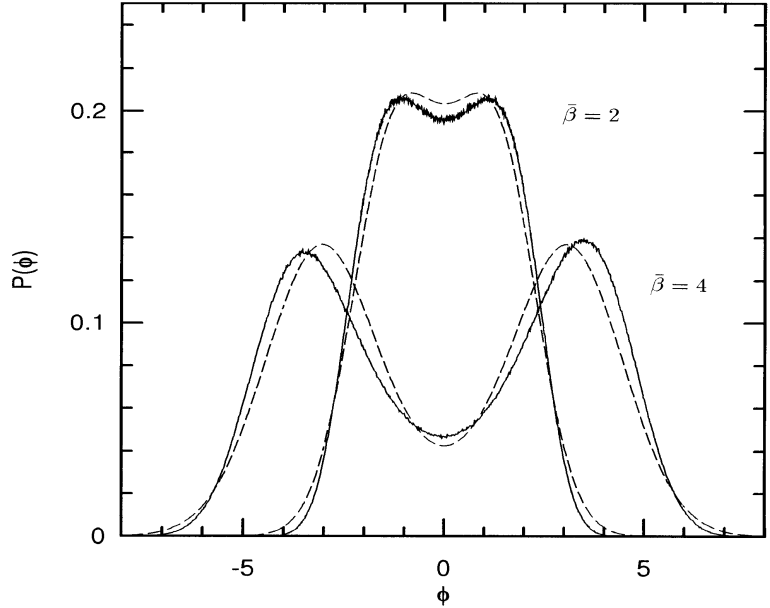,height=8cm,width=10cm}}
\caption[Figure 4]{\small{The classical distribution function $P[\f]$
from the simulation compared with the distribution $\J^2$ from the
double Gaussian approximation (dashed lines). The underestimation of
the wave function overlap in this approximation is already visible at
$\bar{\beta}=4$.}} 
\end{figure}

The equilibrium kink number has traditionally been calculated only in
the phenomenological approach. In Ref. \cite{AHK} an effective kink
number was defined starting from the transfer operator formalism. In
the WKB limit the formula agrees with the standard result up to a
prefactor (which is actually not known explicitly in the standard
calculation). At intermediate temperatures the double Gaussian
approximation can be used to estimate the kink number expected in the
simulations. The results are consistent with the simulations but show
a dependence on an averaging length which again points to the problem
of defining precisely what is a kink at elevated temperatures (Fig.
2). 

The correlation length as computed in the double Gaussian
approximation is in excellent agreement with the simulations for
$\bar{\beta}<5$. There is a smooth crossover in the numerical results
to the exponential dependence on $\bar{\beta}$ predicted by the dilute
gas theory from $\bar{\beta}\sim 6$ onwards. Thus the two theories
agree with the simulations in the appropriate temperature ranges.

Finally, in the double Gaussian approximation one can compute the
internal energy and the specific heat \cite{AHK}. The Schottky anomaly
in the specific heat is clearly seen at $\bar{\b}\simeq 5.4$.
Disentangling the overlap contribution from the diagonal contribution
allows one to associate the peak in $C_v$ with the overlap
contribution, {\em i.e.}, with the kinks (Fig. 5). A similar
computation for the internal energy shows that in this temperature
range the kink contribution dominates over both linear and nonlinear
phonons (Fig. 6). Thus the kinks are indeed relevant to the
thermodynamics of the system at least over some finite range of
temperatures (at low temperatures the thermodynamics is dominated by
linear phonons).

\begin{figure}
\centerline{\psfig{figure=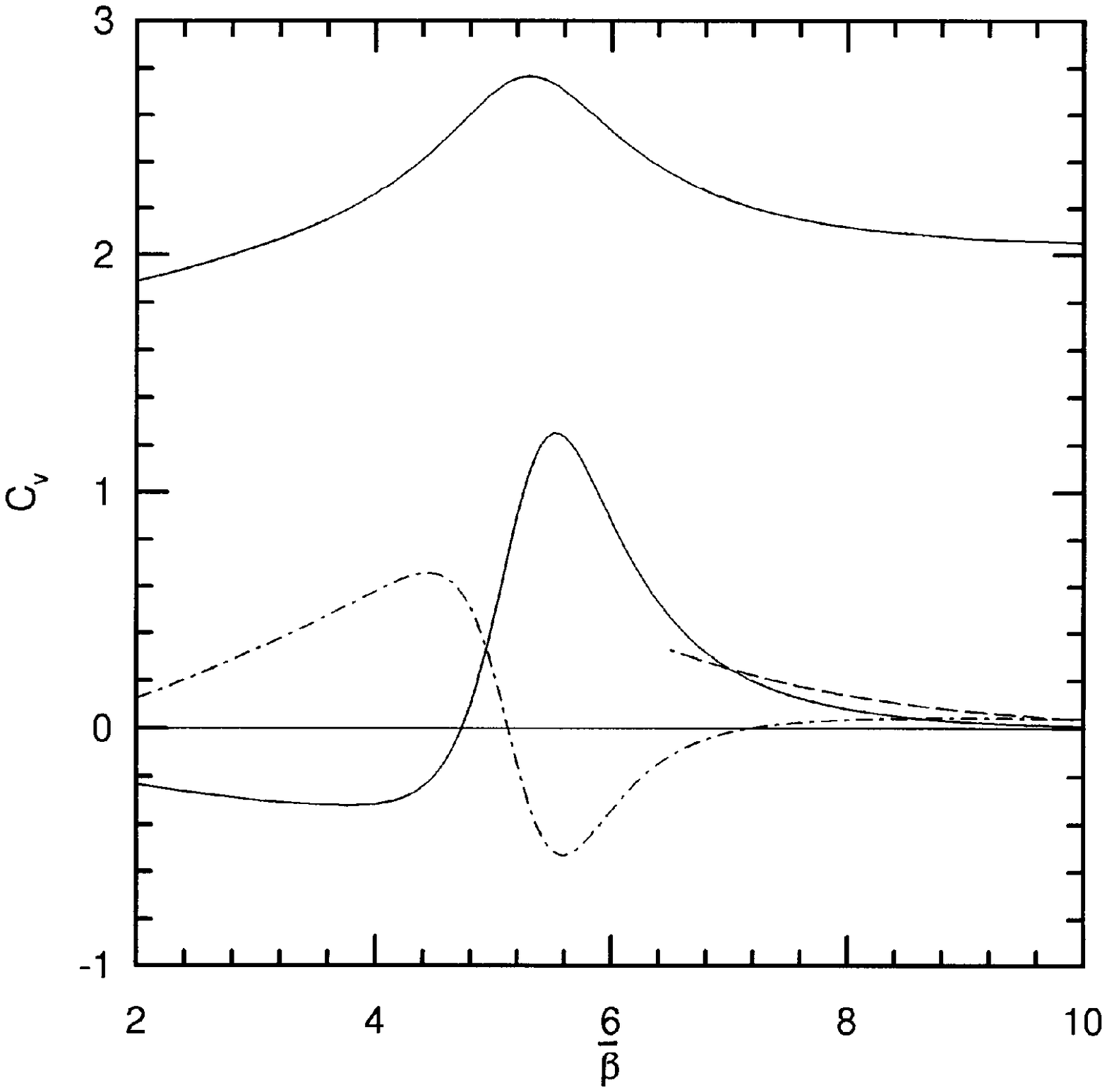,height=8cm,width=10cm}}
\caption[Figure 5]{\small{The specific heat $C_v$ from the double
Gaussian approximation (top curve). The peak in $C_v$ is at
$\bar{\beta}\simeq 5.4$. The contribution of the linear phonons is a
constant (here, this value $=2$). Individual contributions from kinks
(solid line) and nonlinear phonons (dot-dashed line) are shown below.
The WKB calculation for the kink contribution is shown from
$\bar{\beta}=6.5$ onwards.}} 

\centerline{\psfig{figure=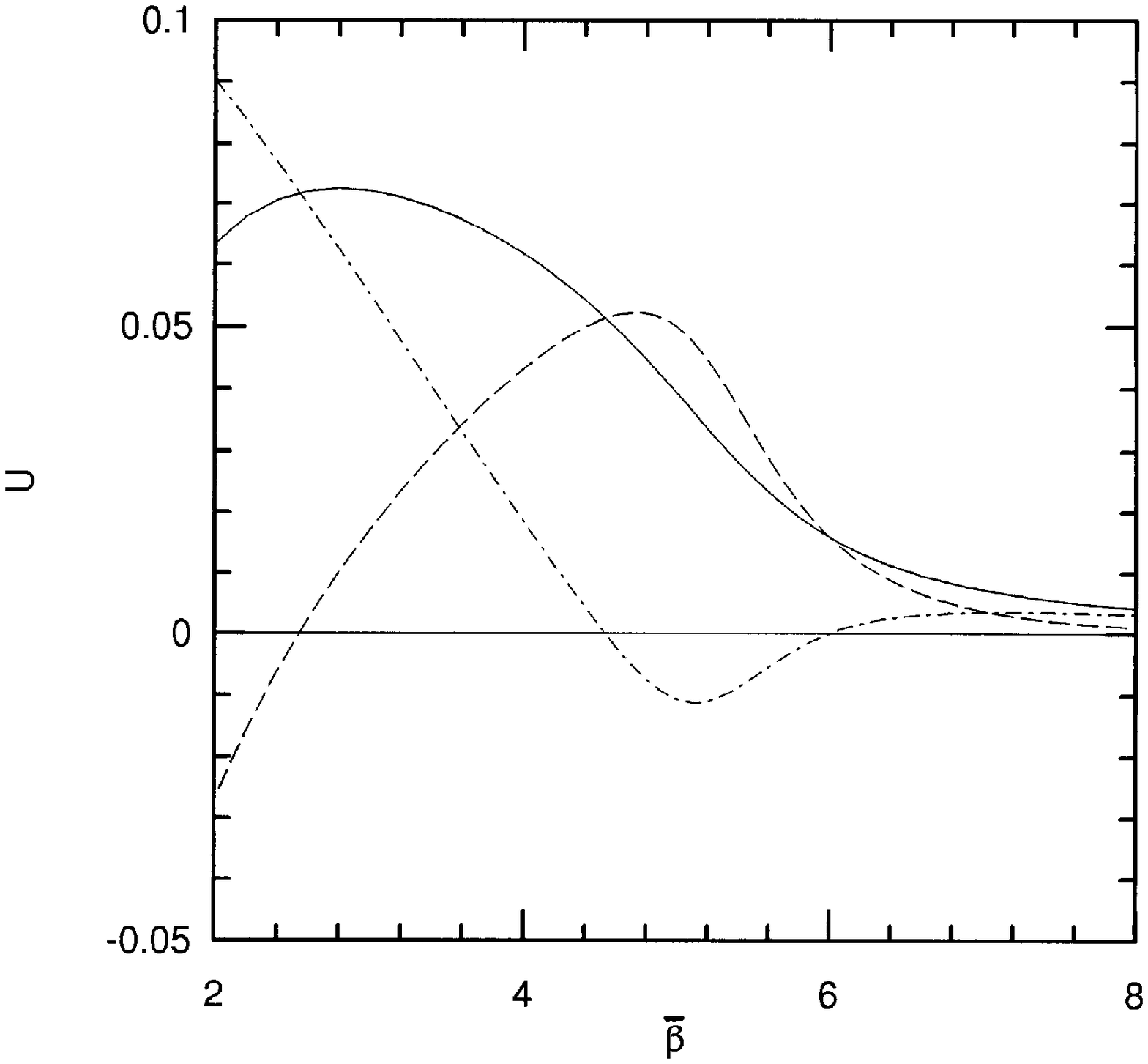,height=8cm,width=10cm}}
\caption[Figure 6]{\small{Internal energy against $\bar{\beta}$ using
the double Gaussian approximation. Kink (dashed line) and nonlinear
phonon (dot-dashed line) contributions are shown separately (modulo an
irrelevant constant energy shift). Note that the kink contribution is
dominant in the region where the specific heat has a maximum.}} 
\end{figure}

\section{Conclusions}

As a consequence of the above results, we conclude that the dilute
gas/WKB approximation is excellent for $\bar{\beta} > 6$ with no
further phonon dressing of the bare kink energy beyond that already
included in (\ref{dgnum}) and (\ref{CL}) at these low temperatures. At
higher temperatures, the WKB analysis fails, though theoretical
progress is possible with the double Gaussian technique. Using this
technique we can analytically calculate the temperature range where
kinks dominate the thermodynamics.

\section{Acknowledgements}

The work reported here was done in collaboration with Frank Alexander
and Alex Kovner. M. Alford, S. Chen, F. Cooper, G. D. Doolen, H.
Feldman, R. Gupta, R. Mainieri, M. Mattis, E. Mottola, W. H. Zurek,
and, especially, A. R. Bishop are thanked for helpful discussions and
encouragement. This research was supported by the U. S. Department of
Energy at Los Alamos National Laboratory and by the Air Force Office
of Scientific Research. Numerical simulations were performed on the
CM-200 at the Advanced Computing Laboratory at Los Alamos National
Laboratory and the CM-2 at the Northeast Parallel Achitecture Center
at Syracuse University.

\end{document}